\documentclass[epj]{svjour}
\usepackage{amsmath,epsfig}
\usepackage{graphicx}
\usepackage{dcolumn}
\usepackage{bm}
\usepackage{amssymb}
\usepackage{amsmath}%
\begin{document}
\title{The scenario of two families of compact stars}
\subtitle{2. Transition from hadronic to quark matter and explosive phenomena}
\author{Alessandro Drago\inst{} \and Giuseppe Pagliara\inst{}}                     
\offprints{}          
\institute{Dip.~di Fisica e Scienze della Terra dell'Universit\`a di Ferrara 
and INFN Sez.~di Ferrara, Via Saragat 1, I-44100 Ferrara, Italy}

\date{Received: date / Revised version: date}
%
\abstract{We will follow the two-families scenario described in the accompanying paper,
in which compact stars having a very small radius and masses not exceeding about 1.5$M_\odot$
are made of hadrons, while more massive compact stars are quark stars. In the present paper
we discuss the dynamics of the transition of a hadronic star into a quark star. We will show that
the transition takes place in two phases: a very rapid one, lasting a few milliseconds, during
which the central region of the star converts into quark matter and the process of conversion
is accelerated by the existence of strong hydrodynamical instabilities, and a second phase,
lasting about ten seconds, during which the process of conversion proceeds as far as the surface of
the star via production and diffusion of strangeness. We will show that these two steps
play a crucial role in the phenomenological implications of the model. We will discuss the 
possible implications of this scenario both for long and for short Gamma Ray Bursts, using the
proto-magnetar model as the reference frame of our discussion. We will show that the process
of quark deconfinement can be connected to specific observed features of the GRBs.
In the case of long GRBs we will discuss the possibility that quark deconfinement
is at the origin of the second peak present in quite a large fraction of bursts. Also we
will discuss the possibility that long GRBs can take place in binary systems without being
associated with a SN explosion.  Concerning short GRBs, quark deconfinement 
can play the crucial role in limiting their duration. Finally we will shortly revisit
the possible relevance of quark deconfinement in some specific type of Supernova explosions,
in particular in the case of very massive progenitors.
\PACS{
      {PACS-key}{describing text of that key}   \and
      {PACS-key}{describing text of that key}
     } 
} 
\maketitle

\section{Introduction}
In the accompanying paper (here and in the following paper 1) we have
discussed the two-families scenario, in which compact stars having a
mass not exceeding about 1.5 $M_\odot$ are made of hadrons, while the
most massive compact stars are entirely made of quarks, i.e. they are
quark stars \cite{Drago:2013fsa}. We have also discussed the
interesting mass range, located about $(1.35-1.5)M_\odot$, which can
be populated both by hadronic and by quark stars. The latters have a
significantly larger radius and a larger moment of inertia. This
scenario is somehow opposite respect to the more traditional one in
which quark stars are extremely compact with radii even smaller than
about 10 km.

In this second paper we discuss how the transition from a hadronic
star into a quark star can take place and which are the
phenomenological implications of that transition.

First, by looking at the plot of the two families already discussed in
the accompanying review paper on the EoS one can recognize the
possible situations in which quark matter and therefore quark stars
can form.  There are essentially three situations:
\begin{itemize}
\item
via mass accretion or via slowing-down of a rapidly rotating pulsar,
having a mass close to the critical one. This situation can for
instance be realized in LMXBs, in which the neutron star accretes mass
and angular momentum from the companion.  Under those conditions the
critical deconfinement density can be reached (maybe more easily soon
after the mass accretion stops and the star starts slowing-down
\cite{Bejger:2011bq}). In these cases the temperature immediately
before deconfinement is quite low and plays no role. We will link this
scenario to the possibility of having long GRBs not connected with a
SN explosion. See also the possible signature of the transition in the
anomalous value of the eccentricity, as discussed in paper 1;
\item
soon after the SN explosion of a massive progenitor. It is possible
that a delay exists between the moment the SN explodes and the moment
quark deconfinement takes place: it can be due again to the gradual
slow-down of the neutron star or to mass accretion due to the
fall-back.  This scenario can be linked to long GRBs displaying two
active periods separated by a quiescent time. 
The possibility that the neutrino flux generated by the phase
transition can help revitalizing a marginally failed SN explosion is not ruled out either. In
this case the temperature before deconfinement can be larger, order of
20-30 MeV. Although these temperatures can facilitate quark
deconfinement they still should not play a crucial role;
\item
after the merging of two neutron stars in a binary system. If a
massive compact star forms immediately after the merging, in our
scheme that star is unavoidably a quark star. This is maybe the most
precise and striking signature of the two-families scenario. Since
these mergers are supposed to be at the origin of short GRBs we expect
to see rather clear signatures of the formation of a quark star in the
features of those GRBs.  Also, the emission in gravitational waves
associated with the process of merging should bear the imprint of the
transition from the hadronic EoS to the quark EoS (see the review
paper by Bauswein et al. of this volume). In the process of merging
very high temperatures are reached (up to about 50 MeV
\cite{Sekiguchi:2011mc} and even larger if the heat released by
quark's deconfinement is taken into account) and we will see that they
play quite a significant role in the phenomenology.
\end{itemize}.

The most important point about the microphysics of the transition from
hadronic matter to quark matter concerns under which conditions the
process of deconfinement can start taking place. The very beginning of
the process is the formation of a droplet of quark matter, stable at
the pressure at which it forms. In the two-families scenario that we
are discussing the formation of quark stars depends on the validity of
the Bodmer-Witten hypothesis on the absolute stability of strange
quark matter \cite{Bodmer:1971we,Witten:1984rs}.  If strange quarks
play a role in the stability of the first droplet than it is clear
that the process of deconfinement cannot start unless some strangeness
content already exists in the hadronic phase (we will come back to
this point when discussing the possible impact of quark deconfinement
on SN explosions).  Statistical fluctuations of the flavor composition
of a small amount of matter can facilitate the formation of an
energetically favorable droplet of quark matter even if the average
strangeness content is not (yet) the optimal one (more strangeness can
form later if the droplet can live long enough that weak interactions
can take place). On the other hand, if the hadronic phase does not
contain any strangeness, either in the form of hyperons or in the form
of condensed kaons, then a droplet of quark matter with a
non-vanishing strangeness content cannot form on the time-scale of
strong interaction, which is the one associated with the fluctuations
of hadrons into deconfined quarks \cite{Iida:1998pi}.  We therefore
assume that the minimal value for the critical density corresponds to
the one at which hyperons (or kaons) start forming. While this density
is well defined at zero temperature, at finite temperature hyperons
form at any density, although their fractional density is very small
at low temperatures and low densities. The exact conditions at which
the first droplet of stable strange quark matter can form at finite
temperature are complicated. A first attempt in exploring that problem
has been made in a few papers
\cite{Bombaci:2009jt,Mintz:2009ay,Bombaci:2011mx}, but it will likely
require more investigations to be completely clarified. We will not
discuss in details the process of formation of the first droplet of
quark matter, because it is analyzed in the papers of Lugones and Bombaci et al. of this volume.

The whole process of quark deconfinement in a stellar object can be divided in different steps:
\begin{itemize}
\item
via quantum fluctuations (if the temperature of the system is low) or thermal fluctuations (if the 
temperature is large) a first droplet of quark matter forms, large enough to keep expanding; 
\item
the droplet keeps expanding (or it merges with other droplets) till its size becomes macroscopic.
This second step has, to our knowledge, never been analyzed; 
\item
the further expansion of the macroscopic bubble of quark matter inside the hadronic star can be described
by using hydrodynamical equations and it divides into two sub-steps \cite{Drago:2015fpa}:
\begin{itemize}
\item
a rapid burning, whose velocity is greatly augmented by hydrodynamical instabilities. It lasts only a few milliseconds
and it burns the central area of the star;
\item
a slow burning, due to production and diffusion of strangeness, lasting some ten seconds and transforming the 
star into a quark star.
\end{itemize}
\end{itemize}

As we will see, the final process of burning can depend on the mass of
the star to be transformed into a quark star and on its initial
temperature. Two sub-cases, at least, need to be discussed: the case
in which the mass of the deconfining star is about (1.4-1.5)$M_\odot$,
the typical situation of deconfinement of a single star via mass
accretion, and the case of deconfinement immediately after the merging
of two compact stars in a binary system, forming a new compact object
with a mass exceeding 2 $M_\odot$. In this second case the temperature
is quite larger and this will have important phenomenological
implications.

\section{Burning of hadronic stars into quark stars}
\label{secburn}
Since the formulation of the Bodmer-Witten hypothesis
 \cite{Bodmer:1971we,Witten:1984rs} and its implication on the
existence of compact stars entirely composed by quark matter
\cite{Alcock:1986hz,Haensel:1986qb}, the process of conversion of
hadronic stars into quark stars has been the subject of many
theoretical investigations. At the microscopic level this process is
extremely complicated because it involves the deconfinement of quarks
(driven by the strong interaction) and flavor changing reactions among
quarks (driven by the weak interaction). In particular the process of
deconfinement is clearly the most complicated due to its
non-perturbative nature. The simple kinetic theory approach proposed
in Ref. \cite{1987PhLB..192...71O} is still one of the most widely
used: the conversion is described as a slow combustion by means of a one
dimensional stationary reaction-diffusion-advection equation for the
strange quarks concentration. The two key quantities in this approach
are the quark diffusion coefficient $D$ ($D \sim 10^{-1}$cm$^2$/sec
for $\mu_q \sim 300$ MeV and $T \sim 10$ MeV \cite{niebergal2010a})
and the time of conversion of down quarks into strange
quarks $\tau$ ($\tau \sim 10^{-9}$ sec for $\mu_q \sim 300$ MeV
\cite{Alford:2014jha}).  By simple dimensional analysis (see
\cite{landau}) one can obtain an estimate of the width of the
combustion zone $\delta \sim \sqrt{D\tau} \sim 10^{-5}$ cm and of the
burning velocity $v\sim \sqrt{D/\tau}\sim10^3-10^4$ cm/sec.

Within the kinetic theory approach of \cite{1987PhLB..192...71O} one
does not take into account possible macroscopic collective flows and
hydrodynamical instabilities driven by pressure and density gradients
between the fuel and the ashes fluids. On the other hand, in the
context of type Ia Supernovae, in which the nuclear burning occurs,
the Rayleigh-Taylor and the Landau-Darrieus instabilities have been
proven to turn the laminar combustion into a much faster turbulent
combustion \cite{Hillebrandt:2000ga,blinnikov}.  In principle one
should couple the equations of hydrodynamics (i.e. the equations of
conservation of baryon number, momentum and energy) and the equation
of conservation of chemical species (which includes the diffusion and
the reaction rates within the combustion zone) in multidimensional
numerical simulations, see \cite{williams}.  Due to the small width of
the combustion zone in comparison with the radius of the star such a
simulation is clearly numerically unfeasible.

In Ref. \cite{Drago:2015fpa} it has been argued that such complicated 
simulations actually are not needed. Indeed one can 
divide the process of conversion 
of a hadronic star into a quark star into two separated regimes:
i) the turbulent regime
which can be described by hydrodynamics under the assumption of an infinitely thin combustion zone;
ii) the diffusive regime in which the two fluids, fuel and ashes are in mechanical equilibrium, but
out of chemical equilibrium. This regime is described by an advection diffusion reaction equation. 
The separation between the two regimes can be found by imposing the so called 
Coll's condition \cite{coll,anile} on the thermodynamical variables of the two fluids 
as we will explain in the following.

\subsection{Turbulent regime}
The turbulent regime can be described within a purely hydrodynamical
approach in which the combustion
zone is so thin to be considered as a surface of discontinuity,
the so called {\it flame front}. We will follow the treatment of
Refs. \cite{coll,anile} where classical
combustion theory has been generalized to the framework of relativistic hydrodynamics.
We indicate with $p_i$,
$e_i$, $n_i$, $w_i=e_i+p_i$ and $X_i=(e_i+p_i)/n_i^2$ the pressure, energy density,
baryon density, enthalpy density and dynamical volume of fluid $i$.
As in the case of the discontinuity associated with a shock
wave, also in the case of the flame front one imposes the
continuity equations for the fluxes of baryon number,
momentum and energy. By indicating with $j$ the number of baryons
ignited per unit time and unit area of the flame front,
the thermodynamical quantities of the hadronic fluid and of the
quark fluid are related to each other by the following equations:
\begin{eqnarray}
n_hu_h&=&n_qu_q=j\\
(p_q-p_h)/(X_h-X_q)&=&j^2\\
\!\!\!w_h(p_h,X_h)X_h-w_q(p_q,X_q)X_q&=&(p_h-p_q)(X_h+X_q)
\end{eqnarray}
the last equation is the so-called relativistic
detonation adiabat. $u_h$ and $u_q$ are the four-velocities of
hadronic and quark matter in the flame front rest frame.  If one starts
from hadronic matter in a initial state A: $p_h=p_A$ and $X_h=X_A$ and with
a given value of $j$, Eqs. 1-3 allow to determine the final state B of
quark matter, $p_q=p_B$ and $X_q=X_B$ which belongs to the detonation
adiabat. The second equation represents a straight line in the (p,X) plane
passing through A and with angular coefficient equal to $-j^2$. The
intersections of this line with the detonation adiabat allow to find
the state B of quark matter. The value of $j$ cannot be expressed in terms of the
thermodynamical variables of the two fluids. It depends in general on
the transport properties of the two fluids (the thermal
conductivity and the diffusion coefficient) and the rate of chemical
reactions. Therefore it must be determined within a kinetic theory
approach such as the one of Ref. \cite{1987PhLB..192...71O}.

\begin{figure}[ptb]
\vskip 0.5cm
\begin{centering}
\epsfig{file=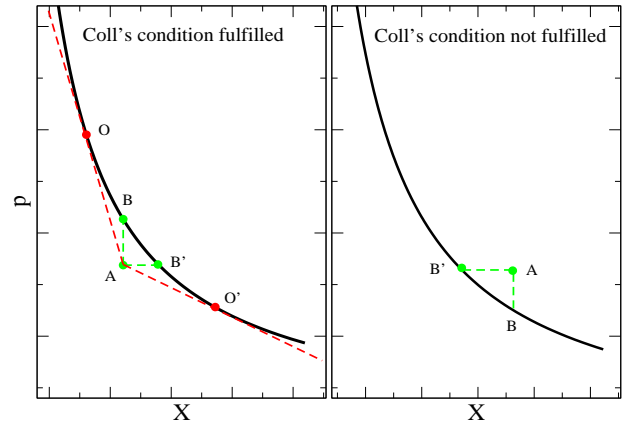,height=8cm,width=5.5cm,angle=-90}
\caption{Illustrative plot of the detonation adiabat in the case in which the Coll's 
condition is fulfilled (left panel) or not (right panel).
A, B, B' indicate  respectively the initial hadronic state and two possible final states for the quark phase. O and O'
are the Chapman-Jouget points. Figure taken from \cite{Drago:2015fpa}.}
\label{pxplane}
\end{centering}
\end{figure}

The so-called
``condition for exothermic combustion'' (``Coll's
condition'') for the conversion of fluid $1$ into fluid $2$ reads:
$e_1(p,X)>e_2(p,X)$, i.e. at fixed pressure and dynamical volume, the
energy density of fluid 1, the fuel, must be larger then the one of
fluid 2, the ash. As shown in Ref. \cite{Drago:2015gsa,Drago:2015fpa} if this condition 
is fulfilled the state A of the hadronic phase lies in the region of the (p,X) plane below the
detonation adiabat (see left panel of Fig. \ref{pxplane}). 
As a consequence, there exist two values of $j$, $j_O$ and $j_{O'}$, for which the lines passing through
A are tangent to the detonation adiabat. The two points of tangency
are the Chapman-Jouget points. In particular, point O corresponds to the
Chapman-Jouget detonation and it is the only possible realization of
detonation in a physical system, such a compact star, in which no
external force is producing the shock wave, see \cite{landau}.
If the Coll's condition is not fulfilled one cannot define the
Chapman-Jouget points and the detonation combustion mode cannot take place.

\begin{figure*}
  \centering
  \begin{minipage}[b]{0.32\linewidth}\centering
    \includegraphics[width=\linewidth]{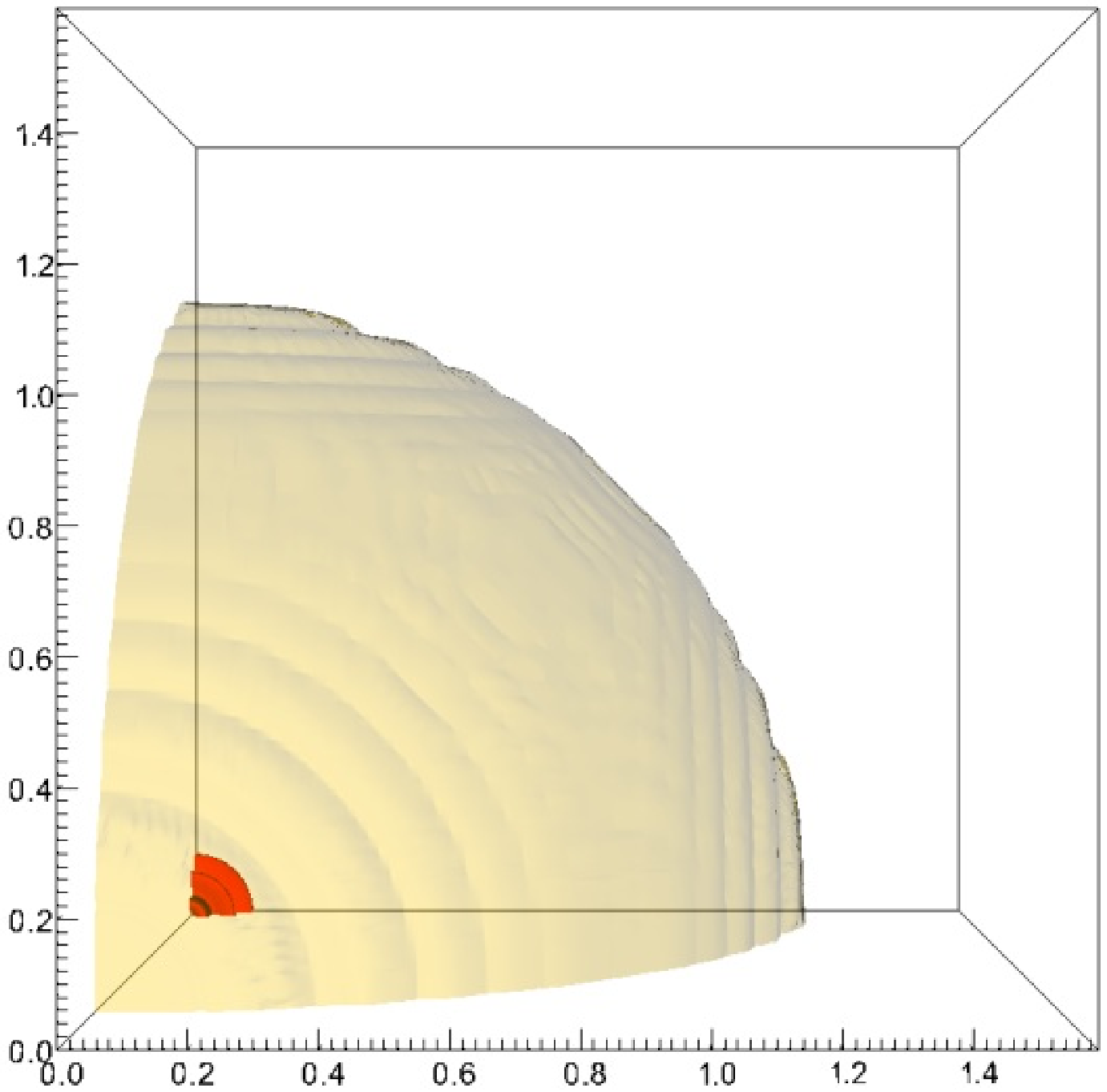}\\
    (a) $t = 0$
 \vspace{1.6cm}
  \end{minipage}
  \hspace{0.08\linewidth}
  \begin{minipage}[b]{0.32\linewidth}\centering
    \includegraphics[width=\linewidth]{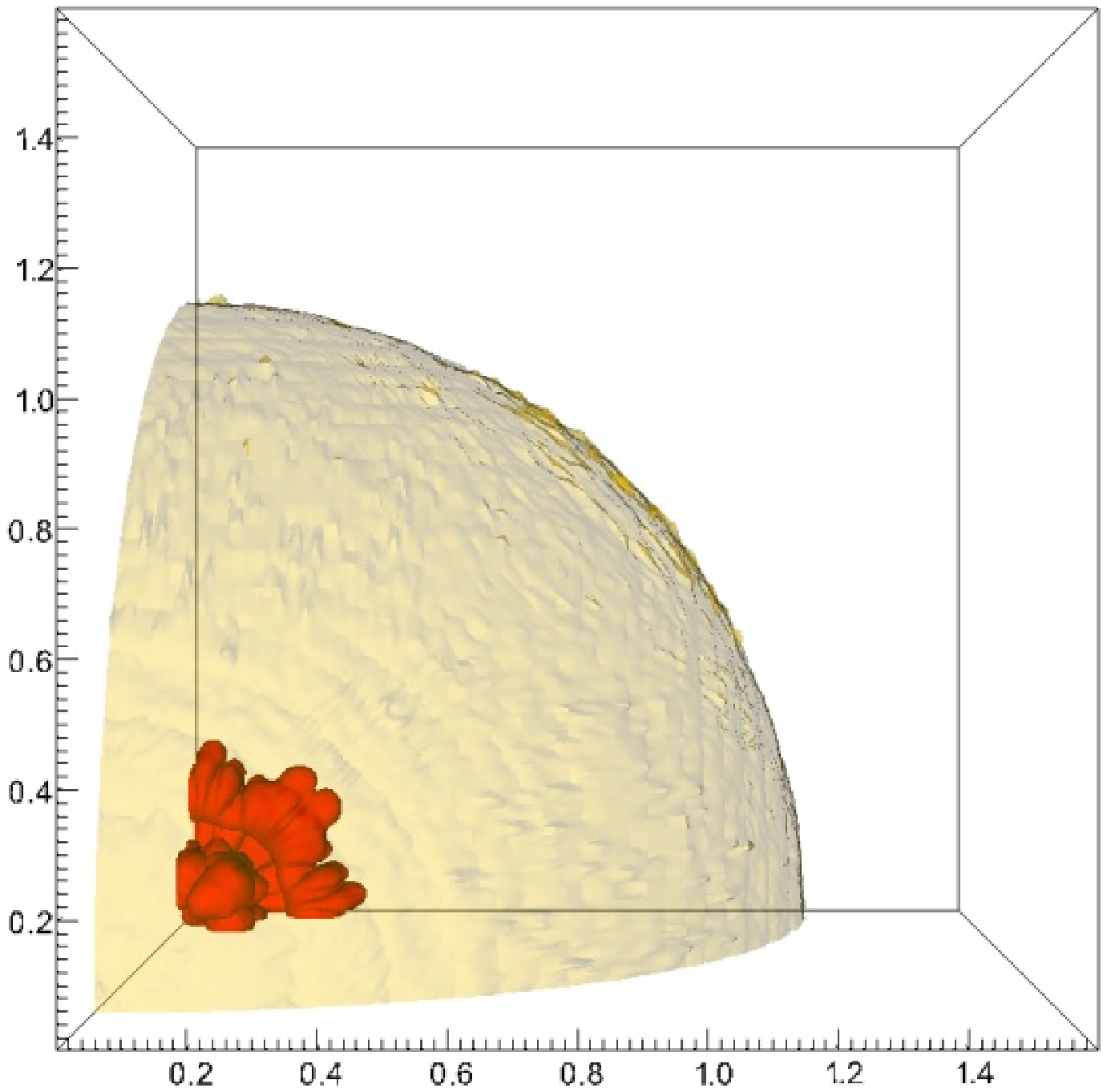}\\
    (b) $t = 0.7\, \mathrm{ms}$
    \vspace{1.6cm}
  \end{minipage}

  \begin{minipage}[b]{0.32\linewidth}\centering
    \includegraphics[width=\linewidth]{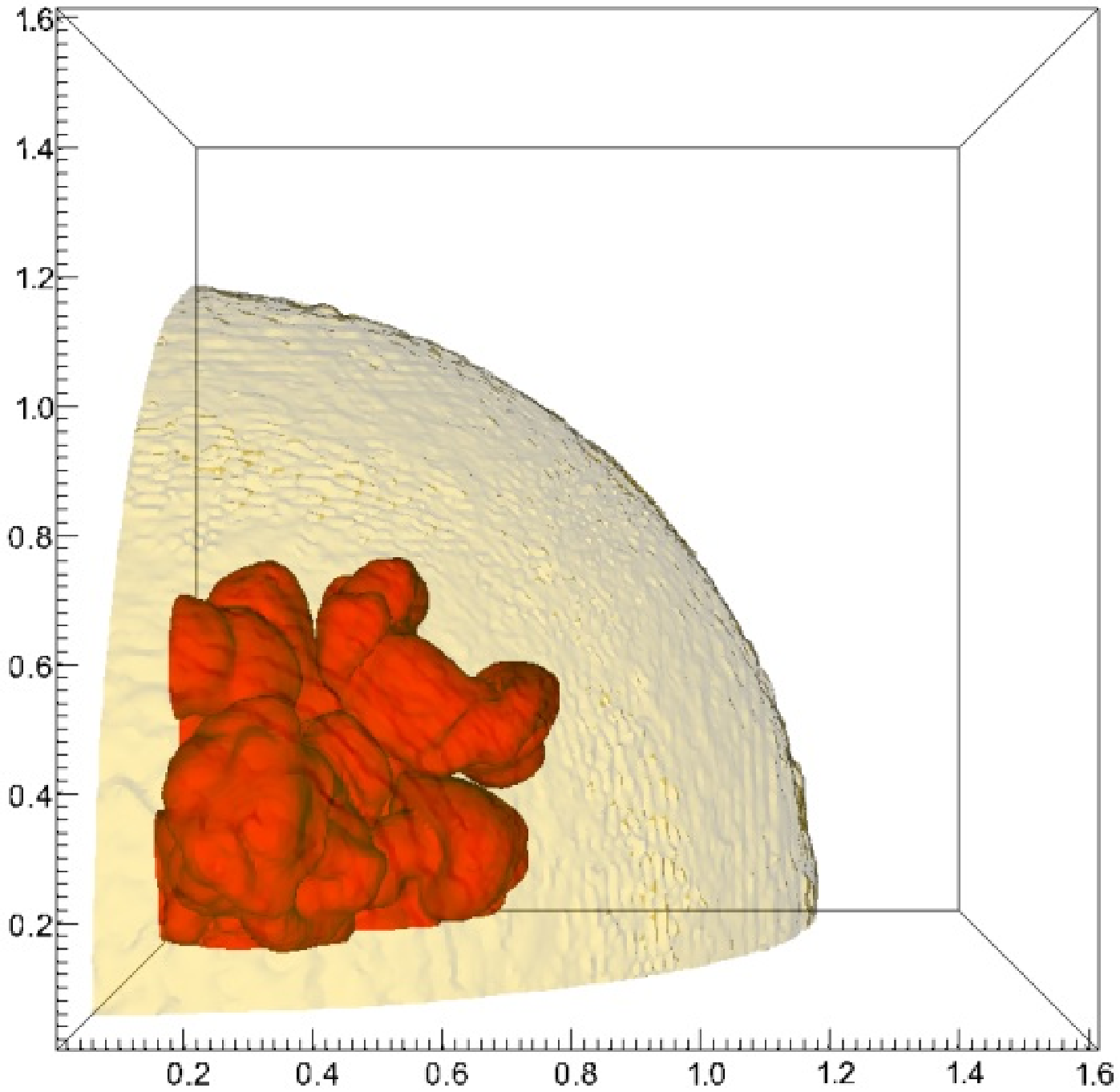}\\
    (c) $t = 1.2\, \mathrm{ms}$
  \end{minipage}
  \hspace{0.08\linewidth}
  \begin{minipage}[b]{0.32\linewidth}\centering
    \includegraphics[width=\linewidth]{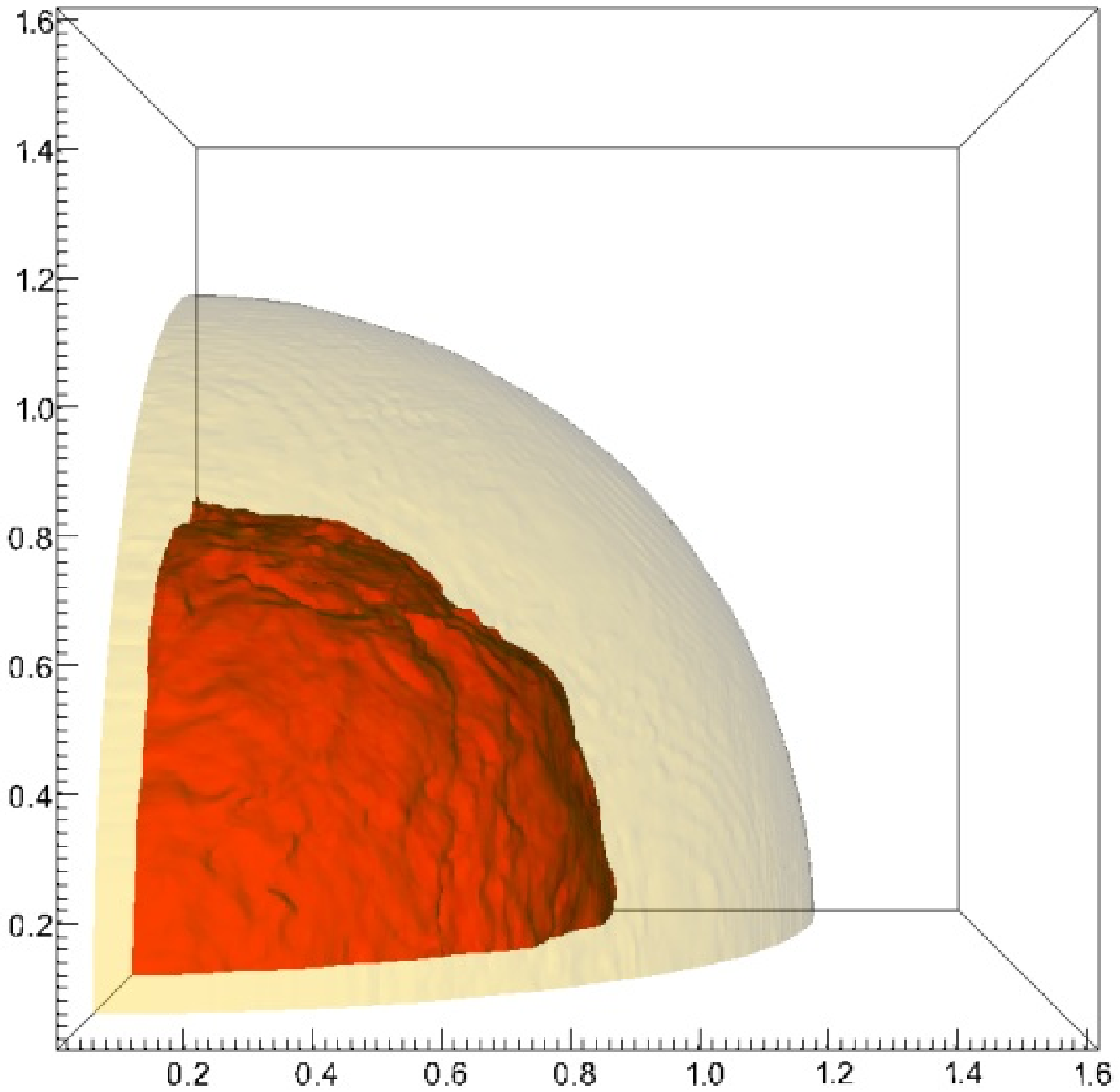}\\
    (d) $t = 4.0\, \mathrm{ms}$
  \end{minipage}

  \caption{(color online) Conversion front (red) and
    surface of the neutron star (yellow) at different times $t$. Spatial units
    $10^6\,\mathrm{cm}$ taken from \cite{Pagliara:2013tza}.}
\label{simulation}
\end{figure*}

Coll's condition is important also to establish whether the
deflagration combustion mode can take place. Let us consider the
simplest case of a slow combustion for which the velocities $v_h$ and
$v_q$ are much smaller than the sound velocities $c_h$ and $c_q$ of
the two fluids.
By using Eqs.(1-3) one finds in this regime that
$p_h=p_A=p_q=p_{B'}$ and $(e_A+p_A)/n_A=(e_{B'}+p_A)/n_{B'}$, i.e. the enthalpy per baryon is
conserved during the combustion (see \cite{landau} for the case of
non-relativistic hydrodynamics). Coll's condition implies that
$X_{B}'>X_A$ i.e.  $(e_{B'}+p_A)/n_{B'}^2 > (e_A+p_A)/n_A^2$ which
together with the conservation of the enthalpy per baryon leads to 
$n_{B'}<n_A$. Moreover, from
$n_A(e_{B'}+p_A)=n_{B'}(e_A+p_A)<n_A(e_A+p_A)$ one obtains
$e_{B'}<e_A$.  Thus the quark phase is produced with baryon density
and energy density smaller than the one of the hadronic phase: 
these conditions are necessary for the Rayleigh-Taylor
instabilities to take place.  
As shown in Refs. \cite{Drago:2005yj,Herzog:2011sn}, the
Rayleigh-Taylor instabilities do indeed occur during the conversion
of a hadronic star and they substantially increase the efficiency of
burning leading to time scales of the order of ms for the conversion
of a big portion of the star \footnote{Notice that our framework is similar to what in the literature
is known as pre-mixed combustion. The
   distinction between a premixed and non-premixed scheme is related to the value of the diffusive burning
   velocity. If that velocity is very large, as suggested in Ref.\cite{niebergal2010a}
   then the increase of the velocity due to
   turbulence is marginal (if any at all). This is close to the result
   of Ref.\cite{Herzog:2011sn} (see Fig.7) because in that paper
   the laminar velocity estimated in \cite{niebergal2010a} has been adopted for the numerical simulations. In our
   scheme, we are instead using the velocities estimated in \cite{1987PhLB..192...71O,Alford:2014jha}
   which are significantly smaller and the
   increase of the velocity due to turbulence is much larger.  This is
   due to the dependence of the Gibson scale on the laminar velocity.}.
In Fig.\ref{simulation}, we display one example of the dynamics 
of the combustion of a hadronic star during the turbulent regime.
The simulation consists in solving
the Euler equations in 3+1D
by using a well-tested grid code that employs a finite
volume discretization, the so-called piecewise parabolic
method, see \cite{Herzog:2011sn} and references therein. Moreover 
a level-set method has been used to follow the evolution of the flame front.
The Rayleigh-Taylor instabilities are clearly visible (the typical mushroom structures)
and render the conversion turbulent.
After about 4 ms, almost the whole star is converted and, at the same time, the
turbulent eddies stop. The star has reached a configuration of 
mechanical equilibrium.
In particular, the pressure, the energy density and
the baryon density of the two phases are continuous at the interface.
This is equivalent to the Coll's equality: $e_h(p,X)=e_q(p,X)$.
The turbulent regime thus stops at the critical density of the hadronic phase, $\overline{n_h}$, for
which the Coll's equality is satisfied.
For $n_h < \overline{n_h}$, the Coll's condition is violated. This implies that the new phase
is produced with $e_{B'}>e_A$ and thus the hydrodynamical instabilities causing 
turbulence cannot anymore take place. 
Notice however that at the interface the temperature and the chemical 
of the two fluids are discontinuous. Therefore
the burning can proceed but with velocities which are
dominated by the diffusion and the rate of the chemical reactions and
which are much smaller than the velocities obtained during
the turbulent regime.  
A natural question arises: is the process still exothermic during the diffusive regime?
As discussed before (see also Ref.\cite{landau}), a slow combustion is characterized 
by the continuity of the pressure and the enthalpy per baryon across the combustion front.
Those two continuity
conditions allow to compute the state of the newly produced quark
matter. We have numerically solved these equations in \cite{Drago:2015fpa} and we have
verified that the new phase is produced at a temperature higher than
the temperature of the fuel.  This is actually the condition of
exothermicity because it implies that some heat will be released from
the star because of the conversion.  It is interesting to notice that
an analytic argument can be provided to show that the conversion
remains exothermic till the surface of the star, see \cite{Drago:2015fpa}. 

\subsection{Diffusive regime}
\begin{figure}[ptb]
\vskip 0.5cm
\begin{centering}
\epsfig{file=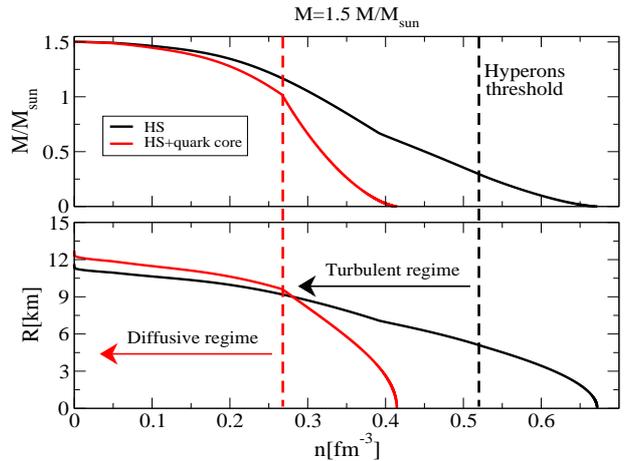,height=8cm,width=6cm,angle=-90}
\caption{Enclosed gravitational mass and radius as a function of the baryon density for a $1.5 M_{\odot}$ 
hadronic star before the turbulent conversion (black lines) and after the turbulent conversion (red lines). The black dashed line marks the appearance 
of hyperons: the seed of strange quark matter is formed
at densities larger than this threshold. The red dashed line marks the density 
below which Coll's condition is no more fulfilled
and the turbulent combustion does not occur anymore. Below this density, the combustion proceeds
via the slow diffusive regime. Figure taken from \cite{Drago:2015fpa}.}
\label{profilo}
\end{centering}
\end{figure}

Let us now discuss how do we model the subsequent evolution of the
conversion during the diffusive regime. First, we need an initial
density profile of the star after the turbulent regime: this
configuration is composed by hot quark matter for densities larger
than $\overline n_h$ and by cold hadronic matter for densities smaller
than $\overline n_h$ (we are discussing here the case of the
conversion of cold hadronic stars). The EoS of hot quark
matter is computed by requiring that at fixed pressure, the enthalpy
per baryon of the quark phase is equal to the one of the hadronic
phase as in the case of a slow combustion. The underlying hypothesis
here is that the kinetic energy of the turbulent eddies taking place
during the first stage of conversion dissipates into heat. Notice that
since the turbulent regime lasts few ms, neutrino cooling (occurring
on time scales of seconds) is not active during the first
stage of the conversion. In Fig. \ref{profilo}, we show one example
for the configuration of a $1.5 M_{\odot}$ hadronic star which
contains hyperons (black lines) which has undergone the turbulent
conversion into a star almost entirely composed by quark matter (red
lines). The upper panel displays the mass enclosed and the lower panel
the radius as functions of the baryon density. Notice that after the
turbulent regime (the density at which this regime stops is indicated
by the red dashed line) a mass of about $0.5 M_{\odot}$ remains
unburnt within a layer with a thickness of about $3$km.

The dynamics of the diffusive regime is regulated by 
two differential equations, one describing the propagation of the
flame front and the other describing the thermal evolution of the star
in presence of the neutrino cooling process and taking into account the heat gradually released 
by the conversion of the layers left unburnt during the turbulent regime.
Concerning the position of the flame front, by labeling 
with $r_f(t)$ its radial coordinate, one can write:
\begin{equation}
\frac{\mathrm{d}r_f}{\mathrm{d}t}=v_{lf}(\mu_q,T)
\label{diff2}
\end{equation}
where $v_{lf}$ is the laminar velocity of the front with respect to
the quark matter fluid (see \cite{Drago:2015fpa}). The initial condition reads
$r_f(0)=\overline{r}$ where $\overline{r}$ is obtained from the baryon
density profile by using the equation
$n_h(\overline{r})=\overline{n_h}$. The thermal evolution is in
principle very complex since one should consider how the heat
progressively generated in the conversion of the external layers is
distributed within the star and one should implement a diffusion
transport code for handling the propagation of neutrinos. This last
task has been treated in  Ref.\cite{Pagliara:2013tza} for the configuration obtained
just after the turbulent regime but without
considering the further conversion of the star in the diffusive
regime. A sensible approximation is to consider the thermal evolution
of the star as being dominated, during the first few seconds, by the
diffusion of the heat deposited during the rapid burning of its
central region (the burning and the cooling of the external layer is sub-leading). 
After this period of time the external layers of the
star are almost isothermal \cite{Pagliara:2013tza} therefore we can make
the simplifying assumption that in the subsequent evolution within the diffusive regime  
the star is basically isothermal.
The simple picture is then the following: the flame propagates towards 
the surface and releases the heat of the conversion; the neutrino cooling operates 
via a black body surface emission with a corresponding luminosity $L=21/8
\sigma (T/K)^4 4 \pi r_s^2$ erg/s \cite{shapiro} with $r_s$ the
radius of the neutrinosphere (we will assume that it is located at the
interface between the inner crust and the outer crust)
The thermal evolution equation then reads:

\begin{equation}
C(T)\frac{\mathrm{d}T}{\mathrm{d}t}=-L(T) + 4\pi r_f^2\, j(r_f,T)\, q(r_f,T)
\label{diff-full}
\end{equation}
where $C$ is the heat capacity of the star, $L$ the neutrino
luminosity, $j$ is the number of baryons ignited per
unit time and unit area and $q$ is the heat per baryon released by the conversion. Concerning the heat capacity, we use $C=2\times
10^{39}M/M_{\odot}(T/10^9)$ erg/K obtained in Ref. \cite{shapiro} for
a uniform density quark star or a hadronic star.  

\begin{figure}[htb]
\vskip 0.5cm
\begin{centering}
\epsfig{file=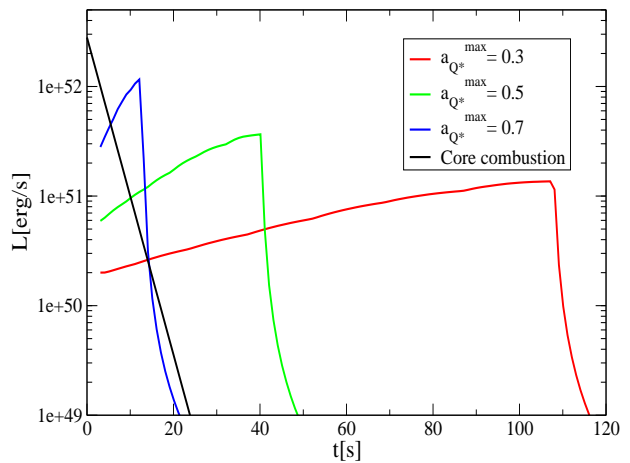,height=8cm,width=6cm,angle=-90}
\caption{Neutrino luminosity associated with the burning during the diffusive regime of the combustion 
for three choices of the parameter $a_{Q*}^{max}$. The black line represents the luminosity obtained from the 
rapid combustion of the core. Figure taken from \cite{Drago:2015fpa}.}
\label{luminosity}
\end{centering}
\end{figure}

By solving simultaneously Eqs. \ref{diff2} and \ref{diff-full} with
initial conditions: $r(0)=\overline{r}$, $T(0)=T_0$ MeV (which is the
temperature of the star for $r>\overline{r}$ after the turbulent
regime and it is of the order of $5$ MeV as found in
\cite{Pagliara:2013tza})
we can calculate the time needed to complete the conversion of the
star and the neutrino luminosity due to the conversion of the material
left unburnt after the turbulent stage. In Fig.\ref{luminosity}, we
show three cases corresponding to different values of the parameter
$a_{Q*}^{max}$ which is related to the minimum amount of strangeness needed 
to render the conversion process exothermic and enters in the expression of $v_{lf}$, see \cite{Drago:2015fpa}.  
We also show a curve of luminosity corresponding to a simple exponential parametrization of the neutrinos released
from the heat generated during the turbulent regime: $L(t)=Q/\tau e^{-t/\tau}$
with $\tau \sim 3$ s and $Q \sim 8.5\times 10^{52}$ erg (see Fig.\ref{luminosity1} for the luminosities computed 
in \cite{Pagliara:2013tza}.)

A remarkable feature is that during the diffusive regime the neutrino
luminosity displays a quasi-plateau (particularly evident for the
smallest value of $a_{Q*}^{max}$). This feature is related to the scaling 
of the  burning velocity with the temperature: $v_{lf} \propto T^{-5/6}$.
As the conversion proceeds, the temperature increases due to the release
of energy and therefore the velocity decreases. It is a
self-regulating mechanism which rapidly leads to an almost constant
velocity of burning and an almost constant luminosity of
neutrinos. The process goes on until the whole star is converted. The
kink appearing in the luminosity curves signals the end of the
conversion: the following evolution is governed only by the cooling
and the standard power law luminosity is obtained.  Typical time
scales to complete the conversion, in this specific case, are of the
order of few tens of seconds. Actually these times scales can be
reduced by considering that, due to gravity, the external
layer will tend to fall onto the conversion front as the flame
propagates. This would lead to an acceleration of the front which
reduces the time of the conversion by roughly a factor of three/four
\footnote{Drago and Pagliara, work in progress.}.

\section{Long GRBs}

GRBs are divided into two subclasses, long GRBs, having a duration of more than 2s, and short
GRBs, lasting less than 2s \cite{Meszaros:2006rc,Berger:2013jza}. 
This division is clearly schematic and one cannot rule out the possibility
that elements of one class intrude the other. The characteristics of the GRBs in the two classes should
derive from the different astrophysical scenario at their origin. 
Short GRBs are generally assumed to be generated by the merging of two compact stars
in a binary system. We will discuss them in the next Section.
The origin of long GRBs instead is typically associated with the
collapse of one massive star, either forming a 
black-hole (collapsar model \cite{MacFadyen:1998vz,MacFadyen:1999mk}) or forming a millisecond
proto-magnetar, a model also known as evolutionary wind model \cite{Metzger:2010pp}. 
At the moment it is not obvious if
all long GRBs should be produced by only one of the two proposed models, or if both possibilities
are realized in Nature, under different initial conditions of the collapsing star. In particular,
the proto-magnetar model requires very strong magnetic fields, of the order of $10^{15}$ G and a rotation period
of the newly formed magnetar of the order of a millisecond. It is not obvious how easy these two 
conditions can be reached and the possibility that, at least in a few cases, some GRB is generated
by the collapsar model is still open, even though long GRBs associated with a SN are probably 
compatible with the proto-magnetar model and not with the collapsar's one \cite{Mazzali:2014iaa}.

In this review we will shortly discuss both possibilities and we will see under which conditions
quark deconfinement can play a role and produce some observable signature.
 
\subsection{Models for long GRBs}
Any realistic model for the inner engine of long GRBs should take into account a few basic
findings:
\begin{itemize}
\item
the total energy emitted in x-rays and in $\gamma$-rays is large, of the order of $10^{50}-10^{51}$ erg;
\item
the typical duration of the initial very luminous phase, called prompt emission,
is of the order of a few tens of seconds, although much longer durations
have been observed in a few cases;
\item
in a significant fraction of long GRBs a prolongated emission has been observed, lasting up to
$10^{3}-10^{4}$ s. While its luminosity is much lower than that of the prompt emission the total energy
emitted during this "quasi-plateau" phase is not much smaller than the energy emitted during the 
prompt phase;
\item
the photons observed during the prompt phase can be well described if one assumes that they are
produced by internal shocks of a ultrarelativistic plasma, expanding with a Lorentz factor $\Gamma$ of the
order of $10^2-10^3$;
\item
the rapid variations in the luminosity of the prompt phase, taking place on a submillisecond scale,
imply that the source has to be compact;
\item
the position of the source has been located in a few cases and it corresponds to a
star formation area of the host-galaxy \cite{Bloom:2000pq}.
\end{itemize}

\begin{figure}[ptb]
\vskip 0.5cm
\begin{centering}
\epsfig{file=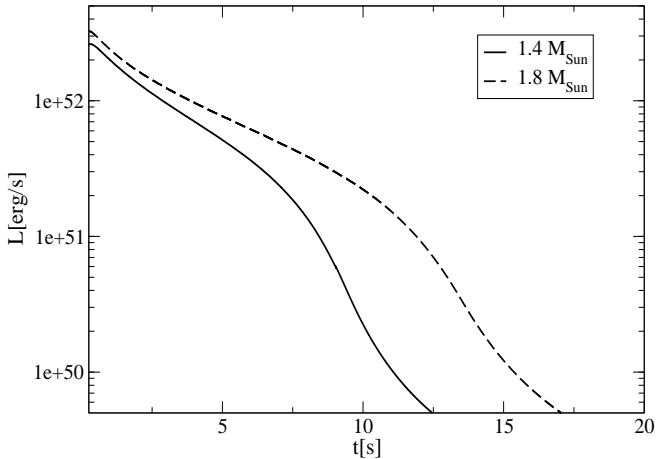,height=8.5cm,width=6cm,angle=-90}
\caption{Total neutrino luminosity as a function of time. The two curves refer to the turbulent (and thus not complete) 
conversion of a $1.4 M_{\odot}$ and a $1.8M_{\odot}$ stars. Figure taken from \cite{Pagliara:2013tza}.}
\label{luminosity1}
\end{centering}
\end{figure}

All these data suggest that the inner engine of long GRBs is a collapsing massive star and that
in many cases
some activity still exists $10^3-10^4$ s after the collapse.

The main difference between the two models lies in the ultimate source of the energy
used to produce the burst: in the collapsar model one uses the energy in the accretion disk
around the black-hole (in principle one can also use the energy of the
rotating black-hole), extracted by the neutrinos. Instead,
in the evolutionary wind model the source of the
energy is the rotational energy of the proto-magnetar. As we will see, in both cases one can imagine
that quark deconfinement can be used to modify the model and the energy associated with the phase
transition can be used to power a burst.
\subsubsection{The collapsar model}
\begin{figure}[ptb]
\vskip 0.5cm
\begin{centering}
\epsfig{file=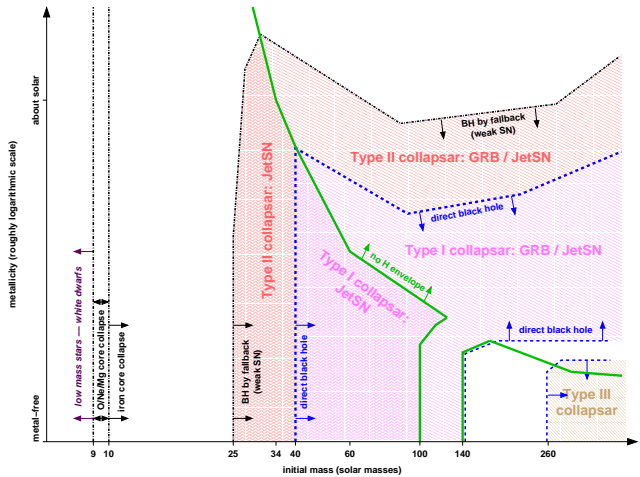,height=9cm,width=7cm,angle=-90}
\caption{Collapsar types resulting from single massive stars as a function of initial metallicity and initial mass. 
The main distinction is between collapsars that form from fallback (Type II; red) and directly (Type I; pink). 
One can subdivide these into those that have a hydrogen envelope (cross hatching), 
only able to form jet-powered supernovae (JetSNe) and hydrogen-free collapsars
(diagonal cross hatching), possibly making either JetSNe or GRBs. The first subclass is
located below the thick green line of loss of the hydrogen envelope and the second is above it. The light
brown diagonal hatching at high mass and low metallicity indicates the regime of very
massive black holes formed directly (Type III collapsars) that collapse on the pair-instability and photo-disintegration. 
Since the collapsars scenario require the formation of a BH, at low mass (left in the figure) or high metallicity (top
of the figure) and in the strip of pair-instability supernovae (lower right) 
no collapsars occur (white). Figure taken from \cite{Heger:2002by}.}
\label{woosley}
\end{centering}
\end{figure}

The central idea behind the collapsar model is rooted into the ultimate fate
of very massive stars \cite{Heger:2002by},
in particular stars having a mass larger than about
$25-30 M_\odot$ and whose external hydrogen and helium layers have been lost
due to strong winds (Wolf-Rayet stars), see Fig. \ref{woosley}. The main points of the model are the following:
\begin{itemize}
\item
the progenitor starts collapsing. A failed supernova follows and a black-hole
forms either directly or due to the large fallback ;
\item
an accretion disk forms. If the angular momentum in the disk is appropriate
most of the energy in the disk can be extracted by neutrino-antineutrino emission \cite{MacFadyen:1998vz};
\item
due to the toroidal geometry neutrino-antineutrino annihilation is a rather efficient 
mechanism and a plasma of electrons and positrons forms in the area around the 
black-hole;
\item
the rotation of the progenitor allows the formation of a empty channel along the 
rotation axis (funnel) on a time scale of the order of ten seconds;
\item
the electron-positron plasma can escape the cocoon of the progenitor along
the funnel. In that way a collimated jet can also form;
\item
a fraction of the energy of the jet can also be used to re-power the supernova
producing a successful explosion of the Ic type.
\end{itemize}
The model is very predictive and this lead initially to spectacular confirmations
and more recently, with more precise data, to some possible problems. In particular
the association between GRBs and Ic SNae has been confirmed in
a few cases \cite{Sollerman:2006ne,Mazzali:2003np,Kulkarni:1998qk}.
On the other hand one problem appeared: the energy of the associated
SN has an energy of about $10^{53}$ erg, much larger than the energy of the jet, what
makes the idea of a SN revitalized by the GRB difficult to justify. The energy of the
SN on the other hand is similar to the rotational energy of a millisecond pulsar,
what can be a strong argument in favor of the proto-magnetar model \cite{Mazzali:2003np}. There is also another
possible problem: at least in one case no associated SN has been observed
\cite{Fynbo:2006mw}.
Finally, in a significant number of cases the prompt emission is made of two well
separated active periods (3 active periods have been observed only in one case),
with a long quiescent time in between. It is still not clear if there is a
statistical evidence of an excess of bursts having long quiescent times respect to the
distribution of all intervals (long or short) separating the active phases. In a few papers
in the past, that evidence was apparently found
\cite{Drago:2005rc,Nakar:2001iz}, but a recent re-analysis \cite{Guidorzi:2015yta} indicates that
maybe all inter-peaks durations can be described by using a same statistical distribution.
On the other hand it is not trivial to explain quiescent times of the order of minutes
just assuming that they are due to some statistical fluctuation: a more detailed description
of how they do take place seems mandatory.

The collapsar model has also some difficulties in explaining the long
emissions taking place, in many cases, after the prompt emission. While the idea
of debris still collapsing onto the black-hole on that long time-scale cannot be
completely ruled-out it seems difficult to justify considering the regularity of the
emission.

\subsubsection{The proto-magnetar model}

\begin{figure*}[ptb]
\vskip 0.5cm
\begin{centering}
\epsfig{file=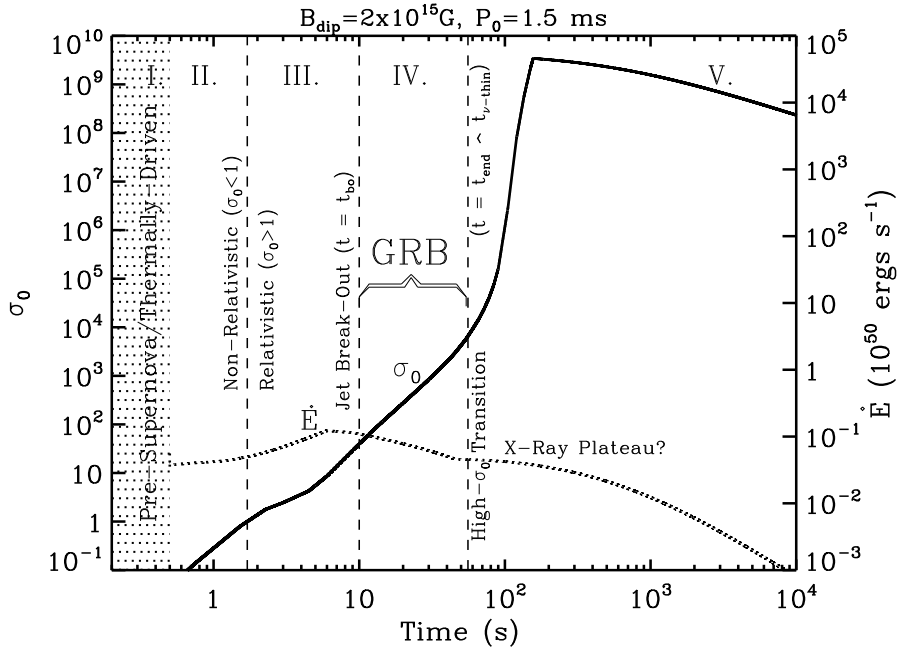,height=8.5cm,width=12cm}
\caption{Wind power $\dot E$
(right axis) and magnetization $\sigma_0$ (left axis) of the proto-magnetar wind as a function of time since
core bounce, calculated for a neutron star with mass $M=1.4 M_{\odot}$, initial spin period $P_0=1.5$ ms, 
surface dipole field strength $B_{dip}=2\times10^{15}$ G, and magnetic obliquity $\chi=\pi/2$. Figure taken 
from \cite{Metzger:2010pp}.}
\label{metzger}
\end{centering}
\end{figure*}

The central idea behind the evolutionary wind model is that at the origin of a long
GRB there is a successful SN producing a rapidly rotating magnetar
\cite{Metzger:2010pp}. The sequence of the
events in the model is the following, see Fig.\ref{metzger}:
\begin{itemize}
\item
after the SN explosion a magnetar forms, with a magnetic field of the order of 
$10^{15}$ G and a period of about one millisecond;
\item
the magnetar starts cooling down by emitting neutrinos and antineutrinos;
\item
whatever charged material is ejected from the star it is strongly accelerated by the 
large Poynting flux $\dot E$ due to the enormous magnetic field and the 
very rapid rotation;
\item
the strong neutrino emission ablates material from the surface of the star.
For some ten seconds the baryon flux is so large that the forming jet has 
a low Lorentz factor due to the baryonic contamination;
\item
the neutrino luminosity reduces and similarly the baryonic flux. The Poynting
flux remains almost constant and a jet with a $\Gamma\sim 10^2-10^3$ can form.
These are the right conditions to generate strong internal shocks in the expanding
plasma: they are at the origin of the observed emission;
\item
the baryonic flux further reduces and the jet becomes almost baryon free.
Under these conditions the motion of the particles in the plasma is almost 
collinear and internal shocks are suppressed. The prompt emission terminates;
\item
the magnetar is still rapidly rotating (although less rapidly) and a pulsar-type
emission can take place, explaining the long quasi-plateau phases
\cite{Lyons:2009ka,Dall'Osso:2010ah}.
\end{itemize}

This model has many interesting features: the most important one is that it allows to explain
the order of magnitude of the energy of the associated SN explosion: it is the rotational energy
of the magnetar. Another extremely nice feature is that it explains in a very natural way
the quasi-plateau emissions, as due to a pulsar-like activity. It is indeed possible to model in a very
precise way all the quasi-plateau just by fitting two numbers: the magnetic field and the rotation's period
\cite{Lyons:2009ka,Dall'Osso:2010ah}.
It has one possible weak point: the maximum Lorentz factor $\sigma_0$ first increases due to the reduction of
the baryonic flux and then slowly decreases, due to the gradual slow-down of the rotation of the star. It is therefore
impossible to explain within the model the 
temporal structure of GRBs having two active periods separated by minutes of quiescence. We will see
how quark deconfinement can solve that problem. Also, the model predicts that all GRBs are associated with
a successful SN, while at least in one case no SN has been observed. Again quark deconfinement can 
provide a possible explanation. 
 
\subsection{The role of quark deconfinement}
As discussed in the previous subsections, there are two problems which are quite difficult
to solve either in the collapsar or in the proto-magnetar model. The first is associated
with GRBs displaying a second peak in the prompt emission, separated from the first
peak by a long quiescent time. The second problem is the possible existence of GRBs not
associated with a SN explosion. In the following we discuss these two problems and we show
how they can be solved in the two-families scenario.
\subsubsection{Long quiescence times}
\begin{figure}[ptb]
\vskip 0.5cm
\begin{centering}
\epsfig{file=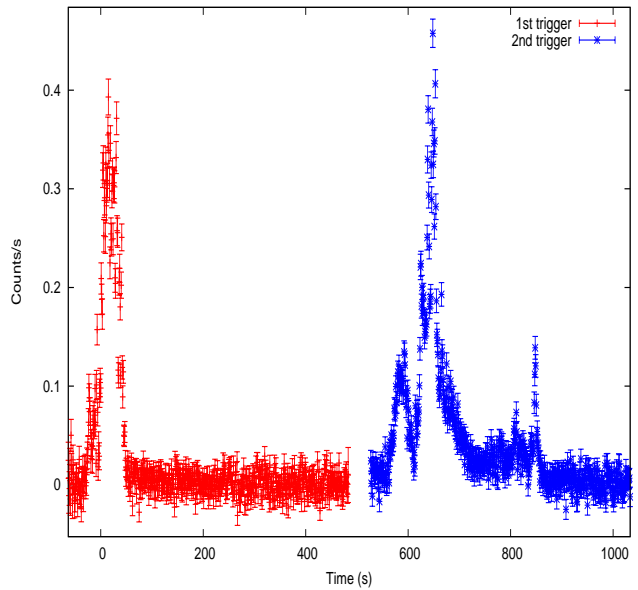,height=8.5cm,width=8cm,angle=-90}
\caption{BAT count rates of GRB 110709B.}
\label{grb1107}
\end{centering}
\end{figure}

In a few cases GRBs display a prompt emission composed of two events
separated by a period of quiescence which can be very long. The most
spectacular case is the one of GRB 110709B \cite{Zhang:2011vk} (see
Fig.\ref{grb1107}) in which the two events are separated by more than
ten minutes. The two emissions present similar luminosities and light
curve characteristics (although with a different time evolution of the
spectral properties).  Interestingly, GRBs presenting more than two
well separated events are very rare and probably the only relevant
example is that of GRB 091024 \cite{Gruber:2011gu} presenting three
episodes of comparable emission. The question is therefore how to
justify what seems a case of reactivation of the inner engine.

The statistical analysis of Refs. \cite{Drago:2005rc,Nakar:2001iz}
indicated an excess of long quiescence times respect to a log-normal
distribution fitted to reproduce inter-pulse durations of whatever
length. Those analysis were therefore suggesting a possible different
physical origin for the long quiescence times. A very recent
reanalysis on the other hand seems to indicate that when the peak
detection efficiency is taken into account the log-normal distribution
has to be substituted with a power-law which is able to describe the
waiting-time distribution of all the pulses. The authors of
\cite{Guidorzi:2015yta} are therefore suggesting that the pulses are
due to the fragmentation of the accretion disk, within the collapsar
model. While that model seems good at interpreting the distribution of
the waiting-times, at least two possibly connected questions remain
open:
\begin{itemize}
\item
the analysis performed in \cite{Drago:2005rc} indicates that on the average the second episode lasts twice as long
as the first one: GRB 110709B is just one representative example of that situation;
\item
explicit simulations are not suggesting that, if the disk fragments, the inner part (the one
powering the first episode) is smaller than the second one.
\end{itemize}

Within the two-families scenario it is rather natural to interpret the second episode as due to
quark deconfinement within the proto-magnetar model. The possible scheme is the following:
\begin{itemize}
\item
the first episode of the long GRB is generated exactly as described with the proto-magnetar model: 
the baryons are ablated from the surface of the compact star by the neutrinos associated
with the cooling of the newly formed compact star whose temperature was of about 20-30 MeV 
immediately after the collapse;
\item
the star starts slowing down (the initial rotation period is of the order of the millisecond) and 
therefore its central density increases;
\item
if the central density of the star during the first episode was slightly below the critical
density needed to deconfine the quarks than during the process of slow-down the critical density
can be reached;
\item
the process of quark deconfinement is strongly exothermic and the inner temperature of the star
increases again up to a temperature comparable to the one reached before;
\item
baryons are again ablated from the surface of the forming quark star, as long as the surface is not
completely converted into quark matter: a new episode of the GRB can therefore take place.
\end{itemize}.

This scheme presents at least a couple of delicate points that will need to be examined in details
in the future.
The first point is the neutrino emission during deconfinement. As we have seen in Section II 
the neutrino luminosity due to deconfinement has two components: one is associated with the
cooling of the central area of the star which has deconfined very rapidly, the other 
is associated with the heat deposited in the outer part of the star while the process of
deconfinement keeps going till it reaches the surface. 
The first component can be approximated as $L_\nu^c\sim Q/\tau_{\mathrm{diff}}\,\, \mathrm{exp}(-t/\tau_{\mathrm{diff}})$.
In the case of long GRBs the heat $Q$ deposited during the rapid deconfinement corresponds roughly to half of the
total deconfinement energy $\Delta E$ of a compact star having a initial mass of about (1.4-1.5$M_\odot$). 
$\Delta E \sim 0.15 M_\odot\sim 4.5 \times 10^{53}$ erg (see paper 1) and therefore
$Q\sim 2 \times 10^{53}$ erg, while $\tau_{\mathrm{diff}}\sim$ (2-3) s. The typical neutrino energy is about 10 MeV.
After some ten seconds, the luminosity of the neutrinos associated with the cooling of the central area
becomes comparable with that associated with the deconfinement of the outer region and, more importantly,
it becomes sufficiently low to allow the possibility of having $\Gamma \sim 10^2-10^3$ if the Poynting
flux $\dot E$ remains similar to the one of the first event.

The second delicate point concerns the evolution of the Poynting flux. One peculiarity of the two-families
model is that the quark star formed after the transition has a larger radius and therefore a larger
moment of inertia than the hadronic star before the transition. Therefore there is a rather strong reduction
of the angular velocity during the transition and this implies a strong reduction of the Poynting
vector unless the magnetic field increases at the same time. The behavior of the magnetic field during
a quark deconfinement phase transition can be quite complicated. Buoyancy forces can
move an internal toroidal magnetic field towards the surface and quark deconfinement can help
by reducing the anti-buoyancy forces \cite{Dai:1998bb,Ruderman:2000np,Heyvaerts:2008bt}. Since the internal magnetic field is typically larger than the
external one it is possible that during the process of quark deconfinement the external magnetic
field increases. In that way the Poynting flux could remain more or less constant. Clearly
at the moment these are little more than speculations and will need to be addressed in future
calculations.

\subsubsection{GRBs not associated with a SN}
Both the collapsar's and the proto-magnetar model are based on a
strict association between a long GRB and a SN explosion. Indeed in a
few cases a type Ic SN has been found associated with the GRB. One can
also imagine that the more the GRB is far away the more it is
difficult to detect the associated SN. On the other hand at least one
case exists, GRB 060614
\cite{Fynbo:2006mw,DellaValle:2006nf,GalYam:2006en}, of a close-by GRB
for which no associated SN has been observed. This suggests the
possibility of GRBs generated through a mechanism not involving a SN
explosion \footnote{It is also possible that the associated SN is sub-luminous because of the fallback
    of $^{56}$Ni onto the BH see Ref.\cite{Moriya:2010cg}. }. The merger of two NSs is such a possibility, but it is
associated with short GRBs, as we will discuss in the next
session. Here we instead consider the possibility of a phase
transition triggered by mass accretion onto the neutron star in a LMXB
system.
In the two families scenario it is rather easy to have quark
deconfinement at the end of mass accretion \cite{Bejger:2011bq} and that transition deposits about
$4.5\times 10^{53}$ erg of heat in the compact star. Still, to
generate a strong GRB one needs to transform that heat into a plasma
made mainly of electron-positron pairs and of photons and to collimate
the jet. One possibility is to consider neutrino-antineutrino
annihilation, a not very efficient process in the case of spherical
symmetry (it is difficult to have head-on collisions between the
neutrinos) so to produce a plasma of about $10^{50}-10^{51}$ erg. The
magnetic field (which in accreting LMXBs cannot be larger than about
$10^{10}$ G) and the rapid rotation would then beam the plasma,
generating a burst that is presumably less energetic and less
collimated than a typical GRB. The ultimate source of the energy of
this burst is the one deposited in the plasma by neutrino-antineutrino
annihilation.

Another possibility is to assume that the neutron star merges with the
white dwarf. It has been shown that this process produces a spinning
Thorne-Zytkow-like object with a low temperature, $T\sim 10^9$ K
\cite{Paschalidis:2011ez}. If large magnetic fields are generated, for
instance via magneto-rotational instabilities, the conditions for
producing a powerful GRB are fulfilled.  Such a burst would be similar
to a short GRB because it is associated with the merger of two compact
stars, but its duration would be comparable with the one of long GRBs.
These features are in agreement with the properties of GRB 060614.

\section{Short GRBs}
Short Gamma-Ray Bursts are characterized by durations typically not
longer than about 2 s, and they are assumed to be associated with the
merger of compact stars (NS-NS or NS-BH) in binary systems
\cite{Berger:2013jza}.

While short GRBs were discovered through their luminous prompt
emission (similarly to what happened in the case of long GRBs) an
extended emission was later found to exist in a significant fraction
of short GRBs \cite{Lazzati:2001gp}.  It was generally assumed that
the prompt emission of short GRBs is spectrally harder than the one of
long GRBs, but the differences are less evident when the sample is
restricted to short GRBs with the highest peak fluxes
\cite{Kaneko:2006ru} or when considering only the first $\sim$2 s of
long GRBs light curves. This was summarized in
Ref.\cite{Ghirlanda:2009de} by saying that when comparing the prompt
emission of short GRBs and the first seconds of long's, one finds: (i)
the same variability, (ii) the same spectrum, (iii) the same
luminosity and (iv) the same $E_{\mathrm{peak}}-L_{\mathrm{iso}}$
correlation.  In other words, if the central engine of a long GRB
would stop after $\sim 0.3\times (1+z)$ seconds the resulting event
would be indistinguishable from a short GRB \cite{Calderone:2014daa}.

\begin{figure}[ptb]
\vskip 0.5cm
\begin{centering}
\epsfig{file=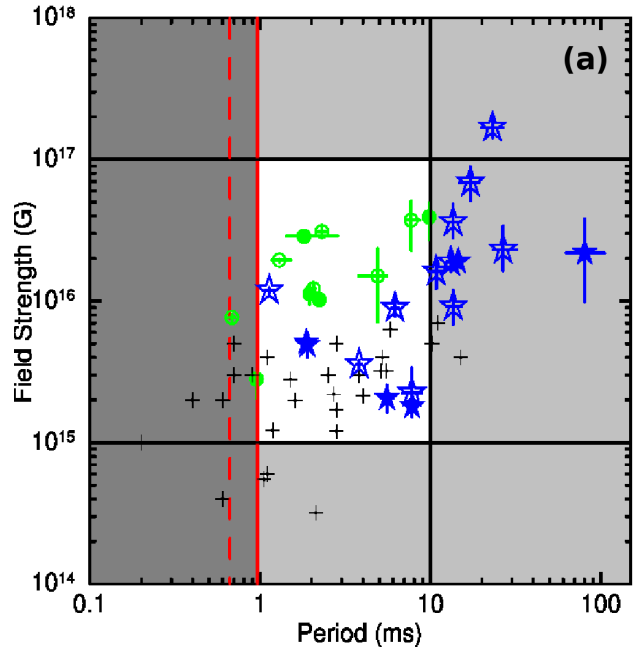,height=8.5cm,width=8cm,angle=0}
\caption{Magnetic field and spin period of the magnetar fits to the extended emission
of both long (black ``+'') and short GRBs. 
The latter are further separated in stable magnetars (blue stars) and unstable 
magnetars collapsing to form a BH (green circles). 
Figure taken from \cite{Rowlinson:2013ue}.}
\label{rowlinson}
\end{centering}
\end{figure}

The similarities between long and short GRBs are not limited to the
prompt emission: actually by comparing the quasi-plateau of long GRBs
and the extended emission of short GRBs one discovers that they can
both be described by assuming that a proto-magnetar formed, rotating
with a period of the order of a few milliseconds and by associating
the prolongated emissions to the pulsar-like emission of that object
\cite{Rowlinson:2013ue}. The rotation period requested is in both
cases of a few milliseconds, the magnetic field is of the order of a
few $10^{15}$G for the long and roughly one order of magnitude larger
for the shorts, see Fig.\ref{rowlinson}.

A recent analysis \cite{Lu:2015rta} suggests that short GRBs can be classified in three categories:
a) those without any extended emission; b) those with an extended emission followed by a rapid
decay of the luminosity; c) those with an extended emission slowly decaying. They propose to
associate the three cases to: a) formation of a BH soon after the merging; b) formation of
a supramassive star collapsing into a BH after having lost part of its angular momentum; c)
formation of a very massive and stable compact star after the merging. In this way they also
estimate the mass distribution of the post-merger remnant as $2.46^{0.13}_{-0.15} M_\odot$.
Although this distribution includes also supramassive stars, it indicates that 
very massive compact stars do exist.


A question naturally arises: if both the long and the short GRBs can be explained, at least in a fraction
of cases, by assuming that a proto-magnetar forms, with similar values for the rotation period
and for the magnetic field, why then the prompt emission of long GRBs lasts tens of seconds and those
of shorts tenths of seconds? In both cases the ablation of material from the surface of the proto-magnetar,
due to neutrino cooling, will provide the crucial ingredient to generate a jet with the proper Lorentz factor.
The cleaner environment and the higher temperatures \cite{Sekiguchi:2011mc} reached after the merger respect
to the post-supernova case would suggest that the duration of the short should be at least comparable
to the one of the long. Which is then the mechanism stopping the prompt emission in the case of short
GRBs? In the next subsection we will show how quark deconfinement can play the crucial role in this
situation.

\subsection{Duration of short GRBs and quark deconfinement}
\label{quarkshort}

One of the best known properties of quark stars is that once formed it is impossible to ablate
hadrons from its surface (unless by neutrinos having energies exceeding about 1 GeV). This is due to 
the confinement of quarks which does not let them to be ejected if not inside a colorless object
as a hadron. A cumulative transfer of energy and momentum to a single quark by multiple
neutrino scattering would also not allow to produce a hadron, because that four-momentum is 
rapidly shared with the other quarks by strong interactions (a similar idea has been discussed in 
Ref.\cite{Paczynski:2005nt}). This property of quark stars
opens the possibility of explaining the rapid truncation of the prompt emission of short
GRBs. Notice that in the two-families scenario, if a compact star (and not a BH)
forms after the merging, it is unavoidably a quark star. 

In Ref.\cite{dlmp:2015} the following scheme has been developed:
\begin{itemize}
\item
a few milliseconds after the merging, the conditions are favourable
for the formation of deconfined quark matter in the hot and rapidly rotating
compact object;
\item
following the scheme described in Sec.\ref{secburn} in a few milliseconds the central
region of the star converts into quark matter. The new object, made of an inner part
of stiff quark matter and of an external part made of hadrons is mechanically stable
(although not chemically stable, since it keeps converting into quarks);
\item
in some ten seconds the star entirely converts into a quark star: until that moment
it still has a surface made of nucleons which can be ablated. Since the star is still
very hot the baryon flux during that first stage is too large to allow the formation
of a jet with a large enough Lorentz factor;
\item
once the surface of the star has completely converted into quarks, baryons can no more
be ablated and the possibility of generating a GRB no more exists;
\item
the GRB can be produced only during the short time associated with the the switch-off
of the baryon flux. The time needed to convert the entire surface of the rapidly rotating
(and therefore strongly deformed star) plays a fundamental role in regulating the
duration of the GRB.
\end{itemize}

\section{Supernovae}
The possible influence of quark deconfinement on SN explosions has been
explored in a few papers, starting from
Refs.\cite{Gentile:1993ma,Drago:1997tn}. In those papers it was
assumed that quark deconfinement takes place before deleptonization
(QDBD following the notation of Ref.\cite{Drago:2008tb}), either at
the moment of the collapse or a fraction of a second after, when
material falls back due to the failure of the SN explosion. 
Since in the
mixed phase the adiabatic index is very low, the collapse continues
rapidly through the mixed phase till the central density reaches the
second critical density separating mixed phase and pure quark matter.
At this point, the adiabatic index becomes large again and the
collapse halts. A shock wave is then produced. One feature of this
mechanism is that it requires a particularly soft EoS, since the
formation of a mixed phase of quarks and hadrons has to take place at
the relatively low densities reached at the moment of core bounce, or
immediately after, during the fallback but anyway before
deleptonization \cite{Sagert:2008ka}. Since the densities reached at
the moment of the bounce are only moderately dependent on the mass of
the progenitor, this mechanism is rather "universal", affecting most
of the SNe, although its effect on the explosion can still depend on
the mass of the progenitor. This first possibility, QDBD, is not
compatible with the two-families scenario, because it would imply that
almost all compact stars are quark stars, since quark deconfinement
would take place at very low densities.

The second possibility, QDAD, is that quark deconfinement takes places
only after an at least partial deleptonization
\cite{Pons:2001ar,Aguilera:2002dh}. It is well known, in fact, that
when the pressure due to leptons decreases, the central baryonic
density increases and therefore the deconfinement process becomes
easier.

The process of deconfinement can then take place in two possible ways:
either as a smooth transition or as a first order one, associated with
the formation of some intermediate meta-stable phase.  In order for
the transition to be completely smooth two conditions need to be
satisfied: finite-volume effects are irrelevant (so that even a single
nucleon can melt into quarks above a given critical density) and no
critical value for the strangeness fraction of the quark phase should
exist.  The latter point is particularly important, because the
existence of a minimum critical value for the strangeness fraction
implies the existence of a second minimum (either local or global) in
the energy per baryon vs density function, separated from the minimum
at strangeness equal zero by a barrier.  In the case of hybrid stars
it is possible to satisfy in particular the second condition, as
analyzed e.g.  in Ref.\cite{Pons:2001ar}. Instead, since the
two-families scenario is based on the existence of quark stars, a
critical value for the strangeness must exist (instead ordinary matter
would just decay into strange quark matter). Therefore the process of
deconfinement in the two-families scenario always goes through the formation of a metastable
phase.

If a critical strangeness fraction needs to be reached in order to deconfine, the question concerns the way
one reaches that critical value soon after the pre-supernova collapse. The most simple way is to imagine that
due to mass fallback and/or to the slow-down in case of a rapidly rotating star the central density increases.
At a certain density hyperons will start being produced. While it is difficult to estimate which can be the
critical density of hyperons (and therefore of strangeness) necessary to trigger quark deconfinement, it is 
clear that it must be at least of the order of a few percent of the total baryons, instead the strange quarks
in the hyperons will be too far away one from the other in order to interact and to drive the process of deconfinement.
As mentioned in the introduction, it is well possible that statistical fluctuations will play the crucial role,
so that at a certain moment a large enough number of hyperons will be contained in a small volume and the
process of deconfinement will start.

A important point can be noticed from Fig. 4 of paper 1: at the temperature reached in the compact star
immediately after the collapse, the densities of hyperons are still not very different from the ones
at zero temperature. Therefore one can conclude that only the stars having a mass close to about
1.5 $M_\odot$ will undergo a phase transition soon after the SN explosion. Stars having a mass of about
1.3-1.4 $M_\odot$ will not be affected and the mechanism by which they explode will not be linked to quark
deconfinement. 

It is well known that at the moment the standard mechanism has
difficulties in explaining SNae associated with the collapse of
massive progenitors, the ones which in principle will generate the
most massive neutron stars. A possible way-out is the following: if
the proto-neutron star is rapidly rotating (a condition similar to the
one needed to produce GRBs in the proto-magnetar model) the rate of
fallback can be reduced, allowing the system the time to deconfine and
to generate a powerful burst of neutrinos associated with the cooling
from the heat released by deconfinement.  Notice that the neutrinos
are generated at a depth of few km inside the star and therefore they
need a few tenths of a second to start flowing out of the star. If the
star is rapidly rotating, so that it does not collapse to a black hole
during that time, the neutrino flux can be sufficient to revitalize
the supernova explosion. In Fig.\ref{luminosity1} we show the neutrino
luminosities computed in Ref.\cite{Pagliara:2013tza} by using a
slightly different set of EoS respect to the ones discussed here: the
luminosities peak at about $3\times 10^{52}$ erg, a value comparable
to the one obtained at the moment of the collapse.  The luminosities
computed with the EoSs described in paper 1 would have even higher
peaks since the total energy released by deconfinement is larger.

\subsection{SN1987a and fizzlers}
The only SN neutrinos detected till now are those from SN1987A.
Indeed, on February 23, 1987, at 2 h 53 m (UT) LSD detector observed 5
events \cite{Aglietta:1987it}; at 7 h 36 m (UT) IMB, Kamiokande-II and
Baksan \cite{Bionta:1987qt,Hirata:1987hu,Alekseev:1988gp} detectors
observed 8, 11 and 5 events respectively.  The progenitor was a blue
supergiant with estimated mass of $\sim$ 20 $M_\odot$.

The observations of Kamiokande-II, IMB and Baksan can be explained
very well within the standard scenario for core collapse SNe, assuming
that the events are due to $\bar\nu_e p\to n e^+$.  The observations
are consistent with the presence of an initial, high luminosity phase
of neutrino emission, followed by a thermal phase due to the cooling
of the newborn neutron star \cite{Loredo:2001rx,Pagliaroli:2008ur}.
Such an initial and luminous phase is expected; indeed, it should
trigger the subsequent explosion of the star.  The standard scenario
for core collapse SNe does not predict the existence of multiple
pulses of neutrino emission and thus cannot accommodate LSD data.

An interesting possibility is that the first burst is due to a very
intense neutronization phase by $e^- p\to n \nu_e$; it was noted in
\cite{Imshennik:2004iya} that electron neutrinos with an energy of
$30-40$ MeV can be more easily seen in the LSD detector than in the
other detectors.  In the astrophysical scenario of
\cite{Imshennik:2004iya}, the rapid rotation of the collapsing core
leads to a delay between the first and the second burst. However, the
nature of the second burst is not discussed in
\cite{Imshennik:2004iya}.

In Ref.\cite{Drago:2008tb} it was discussed the possibility that KamiokandeII, Baksan and IMB
observations are due to the burning of hadrons into quarks.
The sequence of events, in that case, could be the following:
\begin{itemize}
\item
the rapid rotation of the collapsing core halts the collapse at subnuclear
densities, forming a so-called "fizzler" \cite{Imamura:2003};
\item
an initial intense phase of neutronization accounts
for the LSD observations as in \cite{Imshennik:2004iya};
\item
the rapid rotation of the core leads to
the formation of a metastable neutron star, that looses
its angular momentum in a time scale of several hours;
\item
the central density of the metastable star becomes large enough that
deconfinement can take place.
Again, the rapid release of energy at the beginning of the last stage
could be sufficient to lead to the explosion of the star.
\end{itemize}
This scheme, although interesting, is strictly based on the possibility
of having very large values of angular momentum in the central region of the
star, what seems at odd with the results of Ref.\cite{Heger:2004qp}. On the other
hand a similar criticism can be applied to the model for GRBs based on the
formation of a millisecond proto-magnetar, a model that is having a great
phenomenological success. We think therefore that the analysis of the
distribution of the angular momentum in the collapsing core cannot yet
be considered concluded. 

\section{Comparison with other models}
The possible connection between quark deconfinement and explosive
astrophysical phenomena has a relatively long story. The papers
discussing this relation have concentrated on some specific
associations. The oldest proposed connections have to do with
suggestions on how deconfinement can help SNae to explode by providing
a soft EoS in the mixed phase, followed by a stiff EoS in the pure
quark matter phase \cite{Gentile:1993ma,Drago:1997tn}, on how GRBs can
be associated with the energy released by the deconfinement
\cite{Dai:1998bb,Bombaci:2000cv} and on how quark stars can help to
generate the GRB by providing a cleaner environment
\cite{Paczynski:1986px,Haensel:1991um}.

The discovery of very massive compact stars has changed the scenario
concerning the possible impact of quark deconfinement on SNae and on
GRBs, since one needs to clarify the composition
of the most massive stars 
before discussing the transition from hadronic to hybrid or
quark stars. As we have tried to clarify in this review, once a
specific proposal for the EoS of matter at high densities has been
formulated, the possible transitions from hadronic to quark (or, in
other schemes, hybrid) stars appear in a natural way and the
phenomenological implications can be outlined rather precisely.

An attempt at systematizing a variety of phenomena into a unique
scheme has been made during many years by Ouyed and collaborators. The
scheme they have developed, named Quark-Nova, is based on the idea
that the process of deconfinement takes place as a detonation and that
therefore quite a significant amount of matter is ejected by a
mechanical shock at the end of the process. The mass ejection can
interact with the material already present in the surroundings of the
compact star and it can originate a variety of phenomena:
nucleosynthesis in neutron-rich ejecta \cite{Jaikumar:2006qx}; GRBs,
both by releasing a huge amount of energy from the surface of the
quark star via photon emission \cite{Ouyed:2001cg} and also by using
the interaction of the ejecta from the Quark-Nova with the ejecta of
the preceding SN in order to generate a late-time x-ray emission
\cite{Ouyed:2001ts}. Also it has been proposed to explain the
long-time duration and the spectral features of SN 2006gy as due to
the interaction of the Quark-Nova ejecta with the ejecta of the
preceding SN \cite{Ouyed:2010td}.

While the suggested associations between explosive phenomena and quark
deconfinement are very interesting, two inter-correlated questions
arise. First, explicit analysis of the process of quark deconfinement
are not indicating a detonation, but a deflagration, as discussed at
the beginning of this review. Second, it would be interesting to see
if the detonation plays really the crucial role or if the mass ejected
e.g. by neutrino ablation in the case of a deflagration can be
sufficient to interpret some of the phenomena as due to quark's
deconfinement but not to the specific Quark-Nova model. A work in that
direction is e.g. \cite{Keranen:2004vj}, indicating a significant
amount of mass ejection during the formation of a quark star due to
the neutrino emission. This result is compatible with the more recent
analysis we made \cite{Drago:2015fpa} and it opens the possibility of
re-discussing some of the phenomena by using a deflagration instead of
a detonation.

\section{Conclusions}
In this and in the accompanying paper 1 we have presented the
two-families scenario and we have discussed the many implications it
has on astrophysics. We have seen in paper 1 that the measure of the
radius of a few compact stars would likely confirm or rule-out the
model. Possible confirmations could also come from the study of LMXBs
which are displaying in a few cases large eccentricity whose origin is
still unknown, and it could be originated by the phase transition to
quark star of the neutron star in the binary system.

The most spectacular implications of the two-families scenario are
though probably connected with explosive phenomena and in particular
with short GRBs. First, a direct outcome of the two-families scenario
is that if a compact star and not a BH forms after the merging than
that object is a quark star.  This very strong implication can be
tested e.g. by studying gravitational waves emitted immediately before
and immediately after the merger (see e.g. the review paper by
Bauswein, Stergioulas and Janka).  Another striking implication of the
formation of a quark star immediately after the merger is the
possibility of explaining both long and short GRBs by using the
proto-magnetar model as described in Sec.\ref{quarkshort}. Notice that
in these two examples one would not generically test the formation of
quark matter inside the compact star, but the formation of a quark
star, and therefore the two-families scenario.

While many aspects of the scenario still need to be worked out, as for
instance the behaviour of the magnetic field during the formation of
the quark star, we are confident that in the near future the scenario
will be tested and therefore confirmed or ruled out by a multitude of
experiments and observations, ranging from the analysis of GW
emission, to the measure of the radii of compact stars, to the
analysis of the emission of GRBs. The possibility of being tested is
ultimately the divide between a theory and a speculation.

\vskip 0.5cm
 
A.D. would like to thank
V. Hislop for the moral support during the preparation of the paper. 
G.P. acknowledges financial support from the Italian Ministry of Research through the program 
“Rita Levi Montalcini”.

%
%
%
%

\end{document}